\documentclass{PoS}

\usepackage{graphicx}
\usepackage[utf8]{inputenc}
\usepackage{amsmath}

\title{Probing nucleon's spin structures with polarized Drell-Yan in the Fermilab SpinQuest experiment}

\ShortTitle{Polarized Drell-Yan experiment at Fermilab, SpinQuest (E1039)}

\author{\speaker{Andrew Chen}$^{1}$, J.C. Peng$^{1}$, C. H. Leung$^{1}$, M. Tian$^{1}$, N. Makins$^{1}$, 
M. Brooks$^{2}$, \mbox{A. Klein$^{2}$}, D. Kleinjan$^{2}$, 
K. Liu$^{2}$(CoSpokesperson), M. McCumber$^{2}$, P. McGaughey$^{2}$, 
J. Miraal-Martinez$^{2}$, C. Da Silva$^{2}$, Sho Uemura$^{2}$, M. Jen$^{2}$, X. Li$^{2}$,
J. Arrington$^{3}$, \mbox{D. Geesaman$^{3}$,}
P. E. Reimer$^{3}$,
C. Brown$^{4}$, R. J. Tesarek$^{4}$, 
S. Sawada$^{5}$, 
W. Lorenzon$^{6}$, \mbox{R. Raymond$^{6}$}, 
K. Slifer$^{7}$, D. Ruth$^{7}$, 
Y. Goto$^{8}$, 
K. Nakano$^{9}$, T.-A. Shibata$^{9}$,
D. Crabb$^{10}$, D. Day$^{10}$, D. Keller$^{10}$(CoSpokesperson), 
O. Rondon$^{10}$, Z. Akbar$^{10}$,
J. Dunne$^{11}$, \mbox{D. Dutta$^{11}$}, L. El Fassi$^{11}$, H. Jiang$^{11}$,
E. Kinney$^{12}$,
N, Doshita$^{13}$, T. Iwata$^{13}$, \mbox{Y. Miyachi$^{13}$,}
M. Daugherity$^{14}$, D. Isenhower$^{14}$, R. Towell$^{14}$, S. Watson$^{14}$, \mbox{N. Rowlands$^{14}$,} Y. Ngenzi$^{14}$,
S. Pate$^{15}$, V. Papavassiliou$^{15}$, H. Yu$^{15}$, F. Hossain$^{15}$

\\
$^{1}$\mbox{University of Illinois at Urbana-Champaign, Champaign, IL 61801;}
$^{2}$\mbox{Los Alamos National Laboratory, Los Alamos, NM 87545;}
$^{3}$\mbox{Argonne National Laboratory, Lemont, IL 60439;}
$^{4}$\mbox{Fermi National Accelerator Laboratory, Batavia, IL 60510;}
$^{5}$\mbox{KEK, Tsukuba, Ibaraki 305-0801, Japan;}
$^{6}$\mbox{University of Michigan, Ann Arbor, MI 48109-1040; }
$^{7}$\mbox{University of New Hampshire, Durham, NH 03824; } \\
$^{8}$\mbox{RIKEN, Wako, Saitama 351-0198, Japan;}
$^{9}$\mbox{Tokyo Institute of Technology, Meguro, Tokyo 152-8550, Japan;}
$^{10}$\mbox{University of Virginia, Charlottesville, VA 22904;}
$^{11}$\mbox{Mississippi State University, Starkville, MS 39762; }
$^{12}$\mbox{University of Colorado, Boulder, CO 80309;}
$^{13}$\mbox{Yamagata University, Yamagata, Yamagata 990-8560, Japan;}
$^{14}$\mbox{Abilene Christian University, Abilene, TX 79601; }
$^{15}$\mbox{New Mexico State University, Las Cruces, NM 88003}

}

\abstract{

Although the proton was discovered about 100 years ago, its spin structure 
still remains a mystery. Recent studies suggest that the orbital angular 
momentum of sea quarks could significantly contribute to the proton's 
spin. The SeaQuest experiment, which recently completed data collection, 
probed the unpolarized light quark sea distributions of the proton 
using the Drell-Yan process. Its successor, the 
SpinQuest (E1039), will access the $\bar{u}$ and $\bar{d}$ 
Sivers functions using polarized NH$_3$ and ND$_3$ targets. A non-zero Sivers 
asymmetry, observed in SpinQuest, would be a strong indication of non-zero 
sea-quark orbital angular momentum. 
The SpinQuest experiment can also probe the sea quark's transversity 
distribution, which is relevant for the determination of proton's tensor 
charge. Recent study 
suggests that sea-quarks might contribute significantly to deuteron's
tensor polarized structure functions.
This can be further probed in SpinQuest using tensor 
polarized ND$_3$ target. The current status and future 
plan of the experiment are presented.

}

\FullConference{23rd International Spin Physics Symposium - SPIN2018 -\\
		10-14 September, 2018\\
		Ferrara, Italy}

\begin{document}

\section{Introduction}
There has been intense theoretical and experimental interest in the subject 
of transverse momentum dependent (TMD) parton distributions of the nucleons 
\cite{Barone} during the last two decades. For many years after the discovery 
of the partonic structure of the nucleons in the deep inelastic scattering 
(DIS), only their unpolarized and helicity distributions were investigated. 
New insights on the nucleon structure and QCD are expected to be revealed 
by the novel TMD distributions. One example is the striking prediction 
\cite{Brodsky} of the process dependence of certain TMDs, which remains to 
be tested conclusively with ongoing and future experiments. The recent 
advance of the lattice QCD, including the prospect of calculating the 
Bjorken-$x$ dependence of the parton distributions and TMDs \cite{Ji}, could 
also lead to a direct comparison between the experimental data and lattice 
calculations.

From the three transverse quantities of the nucleon and quark, namely, 
the nucleon's transverse spin, quark's transverse spin, and the quark's 
transverse momentum, three different correlations corresponding to three 
different TMDs can be formed. The Sivers function results from the correlation 
between quark's transverse momentum and nucleon's transverse spin \cite{Sivers}. 
The correlation between quark's and nucleon's transverse spins leads to the 
``transversity'' distribution \cite{Barone}. 
The correlation between quark's transverse spin and quark's transverse momentum 
gives the Boer-Mulders function \cite{Boer}. Among the various types of TMDs, 
the bulk of recent interest and progress centered on these three TMDs.

The novel TMDs are accessible via experiments using either lepton or hadron 
beams. Most of the experimental information on the TMDs thus far has been 
collected from the semi-inclusive DIS (SIDIS) experiments carried out with 
muon or electron beams at DESY (HERMES), CERN (COMPASS), and the Jefferson Lab. 
Another important experimental tool to access the TMDs using hadron beams is 
the Drell-Yan process, which involves quark-antiquark annihilating into a 
virtual photon via electromagnetic interaction. As such, the Drell-Yan cross 
sections with unpolarized or transversely polarized nucleons are potentially 
sensitive to various TMDs of the colliding hadrons. First results on the 
Boer-Mulders functions have been extracted from the unpolarized Drell-Yan 
experiments with pion beams \cite{NA10,E615} and proton beam 
\cite{NuSea2007, NuSea2009}. Very recently, Sivers function was probed in 
W-boson production (a generalized Drell-Yan process) using transversely 
polarized proton beam at RHIC by the STAR Collaboration \cite{STAR2016} 
as well as in pion-induced Drell-Yan on a transversely polarized proton 
target at CERN by the COMPASS Collaboration \cite{COMPASS}.

In this paper, we present the physics motivation and experimental status 
of a recently approved polarized Drell-Yan experiment, SpinQuest (E1039), 
at Fermilab \cite{E1039Prop}. The SpinQuest experiment is a follow-up of 
the recently completed SeaQuest experiment \cite{E906Prop}, with an 
important new feature that transversely polarized targets will be implemented. 
The main physics goal of the SpinQuest is to measure proton's sea-quark 
Sivers function, which is expected to provide an unique information 
complementary to that obtained from the polarized Drell-Yan experiments 
at COMPASS and STAR. Additional physics opportunities of the SpinQuest, 
including the measurements of the nucleon's sea-quark transversity 
distribution and the tensor polarized structure function, $b_1$, of the 
deuteron, will also be discussed.  

\section{Probing Sea-quark Sivers Functions}
Sivers first suggested that the correlation between the transverse spin 
of the nucleon and the transverse momentum of the quark (antiquark) can 
lead to single-spin asymmetry in various processes \cite{Sivers}. The 
Sivers function is a time-reversal odd object. Its existence depends on 
initial or final state interactions via soft gluons \cite{Brodsky}. As 
shown in Refs. \cite{Brodsky, HERMES2005}, such interactions are incorporated 
naturally by the gauge link required for a gauge-invariant definition 
of the TMD. The Sivers function is related to the forward scattering 
amplitude where the helicity of the nucleon spin is flipped. This helicity 
flip must involve the orbital angular momentum of the quark, hence the 
existence of the Sivers function indicates non-zero orbital angular 
momentum carried by the quark.

Measurements of the Sivers functions via the polarized SIDIS process with 
transversely polarized targets have been performed at 
HERMES \cite{HERMES2005, HERMES2009}, 
COMPASS \cite{COMPASS2005, COMPASS2015, COMPASS2017} and 
JLab \cite{JLabHallA2011}. 
Both valence and sea quark Sivers functions were extracted via global 
fits \cite{Anselmino, SunYuan} to existing data. The result confirmed the 
theoretical expectation of opposite sign for the $u$ and $d$ valence quark 
Sivers functions. Non-zero values, albeit with large uncertainties, were 
also obtained for the sea-quark Sivers functions \cite{Anselmino, SunYuan}.

The connection between the Sivers function and the orbital angular momentum 
suggests that the Sivers function for antiquarks could be sizable. First of 
all, the meson-cloud model, which describes well \cite{ChengPeng} the 
observed $\bar{d}$/$\bar{u}$ asymmetry \cite{NuSea1998}, implies that a 
significant fraction of nucleon's spin resides in the orbiting meson. Since 
mesons contain valence antiquarks, it is natural to expect that non-zero 
orbital angular momentum is carried by antiquarks. Secondly, using the 
chiral-quark soliton model, it was shown that antiquarks have a dominant 
contribution to proton's orbital angular momentum \cite{Wakamatsu}. Finally, 
recent lattice calculation \cite{Deka} also found a significant fraction of 
proton's spin originating from the $\bar{u}$ and $\bar{d}$ orbital angular 
momentum. All these theoretical results suggest 
that the sea-quark Sivers functions should be sizable and measurable. 
Proton-induced Drell-Yan experiment with transversely polarized target, 
such as the Fermilab SpinQuest experiment, provides a clean probe to extract 
sea-quark Sivers functions. Unlike the polarized SIDIS, where the effects of 
the sea-quark Sivers functions are often overshadowed by that of the 
valence-quark Sivers functions, the polarized Drell-Yan with proton beam is 
predominantly sensitive to the sea-quark Sivers functions.  

\begin{figure}[ht]
\begin{center}
\includegraphics[width=0.5\textwidth ]{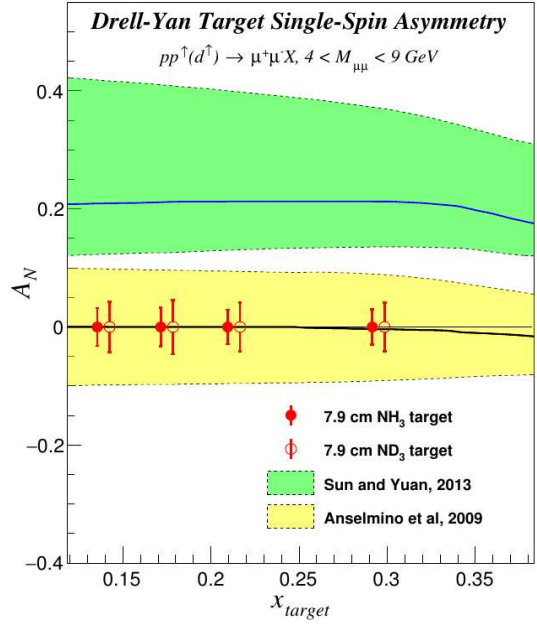}
\caption{Projected sensitivities of the SpinQuest for the $A_N$ measurements 
on NH$_3$ target. The two bands are calculations using the Sivers 
functions obtained from two different global fits to the SIDIS data 
\cite{Anselmino, SunYuan}. }
\label{ProjectionE1039}
\end{center}
\end{figure}

To measure the sea-quark Sivers functions, SpinQuest will use 120 GeV proton 
beam bombarding on transversely polarized NH$_3$ or ND$_3$ targets. Using the 
upgraded SeaQuest dimuon spectrometer, the left-right azimuthal asymmetry,
$A_N$, of dimuons due to the sea-quark Sivers functions in the target nucleon 
will be measured. This asymmetry is expressed as

\begin{equation}
A_N \propto \frac{\sum_q e^2_q [ f_1^q (x_1) {\cdot} {\color{red} f_{1T}^{{\perp}, \bar{q}} (x_2)} + 1 \leftrightarrow 2 ]} 
{\sum_q e^2_q[f^q_1(x_1) {\cdot} {\color{blue} f_1^{\bar{q}}(x_2)} + 1 \leftrightarrow 2 ]} ,
\end{equation}

\noindent where $f_1^q$ and $f^{\perp}_{1T}$ denotes the unpolarized PDF and 
the Sivers function, respectively. The second term in both the numerator and 
denominator has negligible contribution due to the kinematic acceptance of 
the SpinQuest spectrometer. Figure \ref{ProjectionE1039} shows the projected 
sensitivities for the left-right asymmetry $A_N$ as a function of $x$ for the 
measurement on NH$_3$ and ND$_3$ targets. 
The green and yellow bands correspond to the calcuation of $A_N$ 
for the NH$_3$ target using the Sivers functions obtained from two global fits
to the SIDIS data \cite{Anselmino, SunYuan}. 

Figure \ref{ProjectionE1039} shows that the projected sensitivities of 
SpinQuest would allow a significant first measurement of the sea-quark 
Sivers function in the Drell-Yan process. A comparison between the NH$_3$ and 
ND$_3$ results could also be sensitive to the flavor dependence of the sea-quark 
Sivers function. If the magnitude of the sea-quark Sivers function is sizable, 
the sign of the sea-quark Sivers function might also be determined. The 
SpinQuest measurement would complement very well the COMPASS \cite{COMPASS} 
and STAR \cite{STAR2016} measurements, which are sensitive to the sign of the 
valence-quark Sivers functions.

\section{Probing the Sea-quark Transversity}
In addition to the primary physics goal of SpinQuest to measure the sea-qaurk 
Sivers function, an interesting potential by-product of SpinQuest is the 
measurement of nucleon's sea-quark transversity distribution. The transversity 
distribution describes the correlation between quark's transverse spin and 
nucleon's transverse spin. As a chiral-odd object, the transversity is not 
accesible in inclusive DIS, but can be extracted in semi-inclusive DIS or 
polarized Drell-Yan experiments, such as SpinQuest \cite{Barone}. In order to 
probe the chiral-odd transversity distribution, another chiral-odd object is 
involved in the measurement. For the SIDIS, the chiral-odd Collins 
fragmentation function \cite{Collins} or the interference fragmentation 
function are involved. For the Drell-Yan process, a convolution of a 
transversity distribution with either another transversity distribution 
or the chiral-odd Boer-Mulders function is involved. 

Extensive efforts to measure the transversity distributions via SIDIS have 
been performed at HERMES \cite{HERMES2005, HERMES2010}, 
COMPASS \cite{COMPASS2007, COMPASS2012}, and 
JLab \cite{JLabHallA2011}. 
A global fit \cite{Anselmino2013} to these SIDIS data, as well as the 
$e^+ e^-$ data from Belle \cite{Belle2006} relevant for the Collins 
fragmentation function, has led to the extraction of the quark transversity 
distributions in the proton. However, in this global analysis, the sea-quark 
transversity distributions were assumed to be zero. Again, this reflects the 
challenge encountered in the SIDIS to separate the antiquark contribution 
from the dominant quark contribution.  

The tensor charge of the proton is defined as the first moment of the 
difference of the quark and antiquark transversity distribution as

\begin{equation}
{ \delta {q} = \int_0^1 [{h_1^{q}(x) - h_1^{\bar{q}}(x) }] {dx} }
\end{equation}

\begin{figure}[ht]
\begin{center}
\includegraphics[width=0.5\textwidth ]{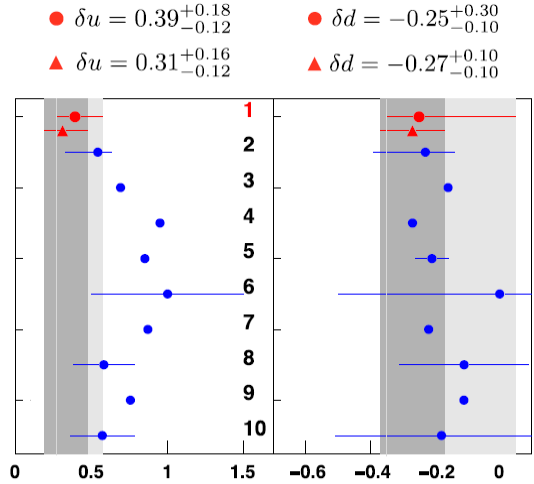}
\caption{Tensor charges for the $u$ and $d$ quarks extracted in the global 
fit to SIDIS data \cite{Anselmino2013}. Predictions from various models and 
the lattice calculations are also shown.}
\label{tensorCharge_global}
\end{center}
\end{figure}

\noindent From the global fit to the SIDIS data, the $u$ and $d$ quark tensor charges 
were obtained \cite{Anselmino2013} as shown in Figure \ref{tensorCharge_global}. 
While the tensor charge for the $d$ quark is in reasonable agreement with 
model calculations, Figure \ref{tensorCharge_global} shows a significant 
discrepancy between the data and model calculations (and lattice calculations) 
for the tensor charge of the $u$ quark. A possible explanation is that the 
sea-quark transversity distributions, which were neglected in the global 
analysis, might be quite sizable. It would be of great interest to measure 
the $\bar{u}$ and $\bar{d}$ transversity distributions using the Drell-Yan 
process.

In the SpinQuest experiment, the antiquark transversity distributions can be 
accessed through the measurement of the correlation between the azimuthal 
angle of the charged leptons from the Drell-Yan process and the transverse 
polarization direction of the target nucleon. Specifically, the Drell-Yan 
differential cross section on a transversely polarized nucleon contains a 
term proportional to sin$(2\phi - \phi_s)$, where $\phi$ is the azimuthal 
angle of charged lepton in the dilepton rest frame and $\phi_s$ is the 
azimuthal angle of the nucleon's spin direction \cite{COMPASS}. The 
amplitude of this term is proportional to the convolution of the quark 
Boer-Mulders function of the proton and the antiquark transversity 
distribution of the target proton. As the Boer-Mulders function can be 
determined from the $\cos 2\phi$ distribution 
of the unpolarized Drell-Yan data \cite{NA10, E615, NuSea2007, NuSea2009}, 
it is quite feasible to probe the sea-quark transversity distributions. 
Further study of the expected sensitivity in SpinQuest for such measurement 
is underway.

\section{Study of Deuteron Tensor Polarization Function}
\begin{figure}[ht]
\begin{center}
\includegraphics[width=0.5\textwidth ]{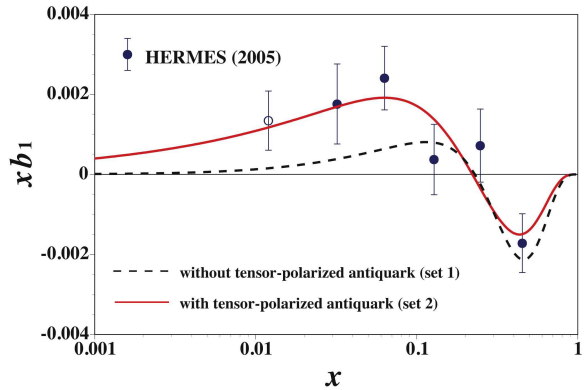}
\caption{
Comparison between the HERMES data \cite{HERMES2005-95} and the theoretical 
prediction \cite{Kumano2014}. The solid and dashed curves correspond to 
calculation with, or without, finite antiquark tensor polarization 
respectively.
}
\label{xb1_HERMES}
\end{center}
\end{figure}

With tensor polarized ND$_3$ target, the SpinQuest experiment can measure the 
tensor polarization function of the deuteron. Deuteron has spin 1 thus has 
three magnetic sub-states, m = +1, 0, -1.  This leads to new PDF functions, 
$b_1$, $b_2$, $b_3$ and $b_4$. Among which $b_1$ is related to the difference 
between |m| = 1 and 0. Figure \ref{xb1_HERMES} shows the HERMES SIDIS 
measurement \cite{HERMES2005-95} of $xb_1$ vs $x$, where $x$ is the 
Bjorken-$x$. The HERMES $b_1$ data suggest an oscillating pattern with a node 
occuring at $x \sim 0.3$, in qualitative agreement with a conventional description 
for $b_1$ \cite{Hoodbhoy1989}. The two curves in Figure \ref{xb1_HERMES} correspond
to theoretical calculations \cite{Kumano2014} for the $xb_1$ distribution. 
The solid curve, which assumes non-zero antiquark contribution to $b_1$, gives 
significantly better description than the calculation (dashed curve) which assumes 
no antiquark $b_1$ contribution. This suggests a sizable antiquark contribution to 
the deuteron's $b_1$ structure function.

\hfill

\begin{figure}[ht]
\begin{center}
\includegraphics[width=0.5\textwidth ]{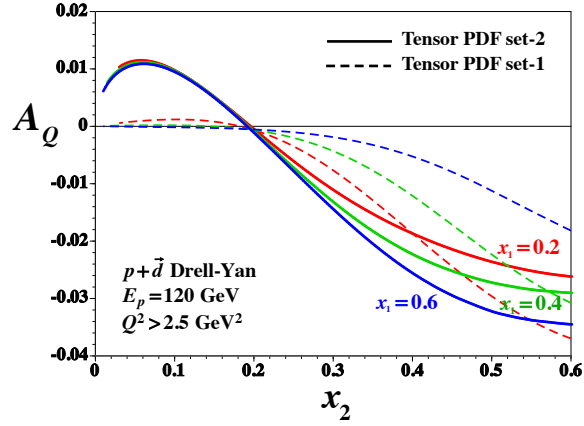}
\caption{A prediction of the asymmetry $A_Q$ for the SpinQuest 
experiment based on two tensor PDF sets obtained from a fit to the HERMES 
data.\cite{Kumano2014} The solid curves were obtained with finite antiquark 
tensor polarization, while the dashed curves correspond to calculations
assuming vanishing antiquark tensor polarization.}
\label{extrap_E1039}
\end{center}
\end{figure}

\begin{figure}[ht]
\begin{center}
\includegraphics[width=0.5\textwidth ]{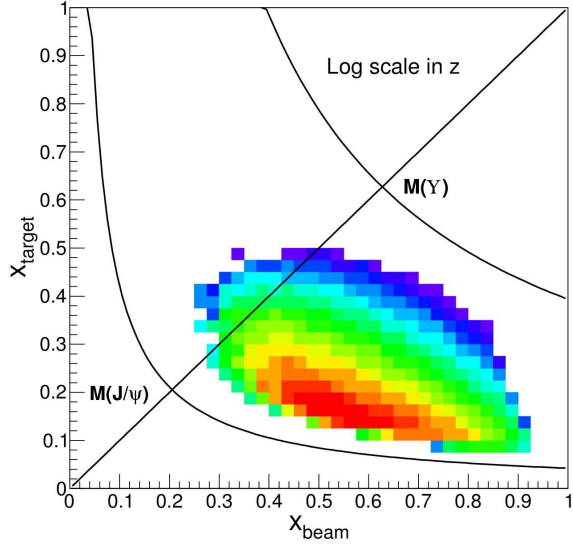}
\caption{The 2D distribution of $x_{target}$ ($x_2$) vs $x_{beam}$ ($x_1$) 
showing that the SpinQuest can cover $x_2$ down to 0.1  
to measure the tensor polarization function.}
\label{x1_x2}
\end{center}
\end{figure}

The SpinQuest experiment can provide a direct measurement of the sea-quark
contribution to the $b_1$ structure function. Using a 
tensor polarized ND$_3$ target, the tensor polarization asymmetry, $A_Q$, 
in the Drell-Yan process can be measured. $A_Q$ is defined as the ratio of 
the unpolarized PDFs and the tensor-polarized PDFs \cite{Kumano2014}.
Figure \ref{extrap_E1039} shows the prediction for $A_Q$ for the SpinQuest
kinematics \cite{Kumano2014}. 
The solid lines are the prediction assuming non-zero antiquark 
tensor polarization PDFs with various values of $x_1$ , the momentum fraction 
of quark inside the beam proton. While the dashed lines assume zero antiquark 
tensor polarization. 
This shows that $A_Q$ can distinguish the two predictions, 
especially at the $x_2 < 0.2$ region. Figure \ref{x1_x2} shows that the
SpinQuest experiment covers the appropriate ranges of $x_1$ and $x_2$ to
measure the $b_1$ of deuteron. Dedicated beam time to take both tensor 
polarized and unpolarized data would be necessary to make a quality 
measurement.

\section{The SpinQuest Experiment}
\begin{figure}[ht]
\begin{center}
\includegraphics[width=1.0\textwidth ]{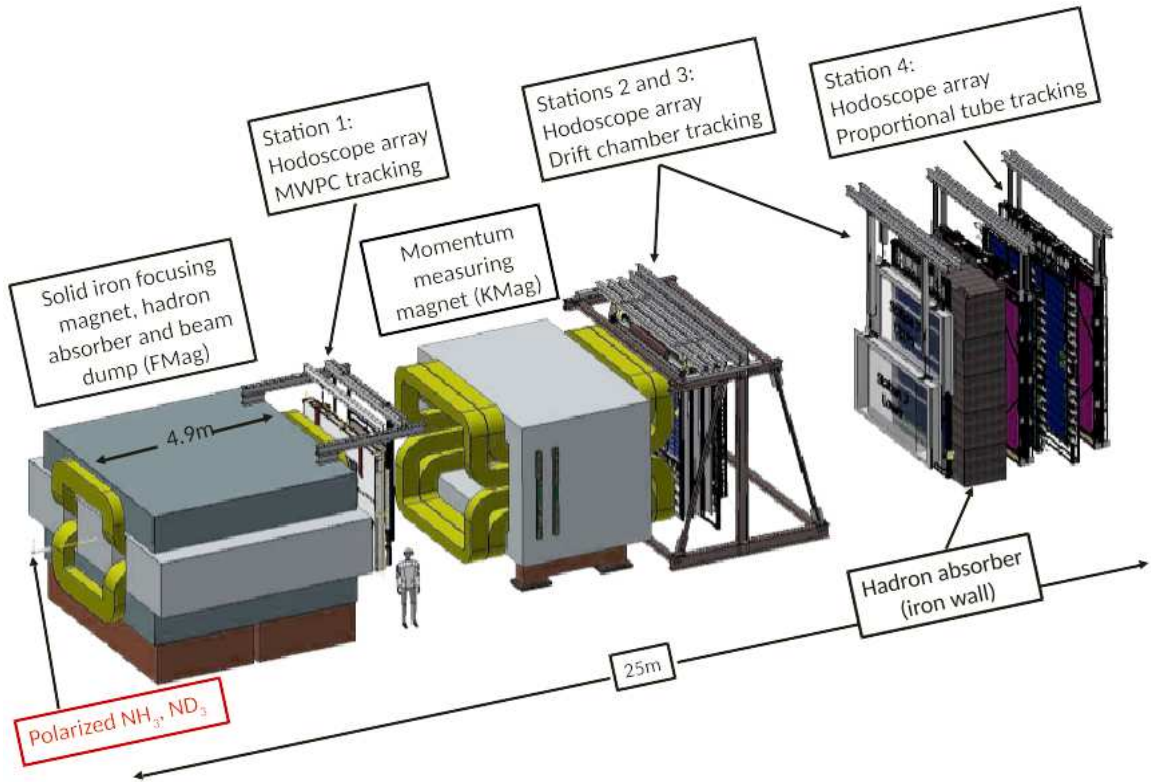}
\caption{The spectrometer of the SeaQuest/E906 is used in SpinQuest experiment.}
\label{E906Spectrometer}
\end{center}
\end{figure}

The SpinQuest spectrometer is basically the same as that used in the 
SeaQuest/E906 experiment \cite{E906Prop}, shown in 
Figure \ref{E906Spectrometer}. 
The liquid H$_2$ and D$_2$ targets, used in the SeaQuest experiment, will
be replaced by the polarized NH$_3$ or ND$_3$ target;
also moved upstream by 1.5 meter to improve the separation of dimuon 
events from target and from dump. New fiber scintillators will be added 
in stations 1 and 2. This provides better $y$ position resolution thus 
improving the $z$ resolution and background rejection. 
The target is followed by a solid iron bending magnet, 
which also servers as a beam dump. 
Dimuons are tracked by stations 1-3 consisting of drift chambers and 
hodoscopes, while a second magnet (KMAG) is for analyzing the muon 
momentum. Station 4, placed downstream of an iron block, consists of
hodoscope and proportional tubes for muon identification. The hit
patterns of the hodoscopes are utilized to select muon events at the
trigger level.

\hfill

\begin{figure}[ht]
\begin{center}
\includegraphics[width=0.8\textwidth ]{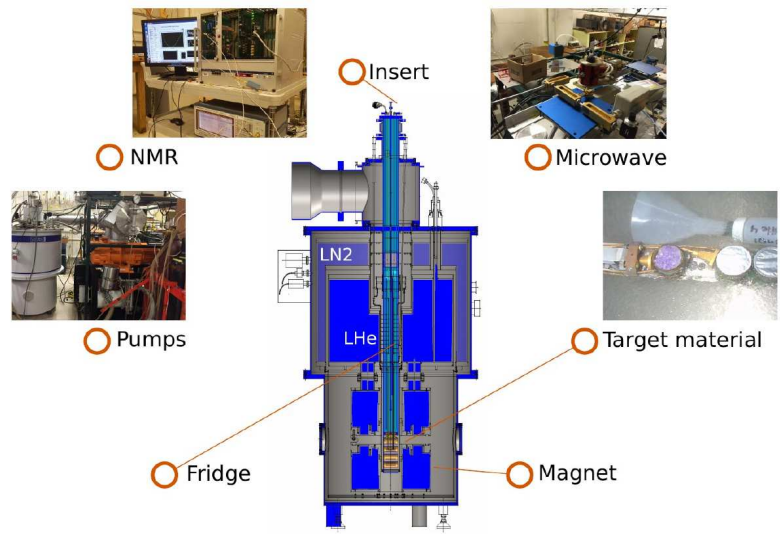}
\caption{The target system of SpinQuest shown together with its 
supporting devices.}
\label{targetE1039}
\end{center}
\end{figure}

Shown in Figure \ref{targetE1039} is the polarized target system of 
the SpinQuest. The target is polarized using Dynamic Nuclear Polarization
as described in \cite{Crabb1995}.
The frozen ammonia beads of NH$_3$ and ND$_3$ will be placed in the target 
holder in a 5 T magnet field. The whole container is cooled down to 
1 K in a He evaporation refrigerator.
In a 5 T field at 1 K, electrons are nearly fully polarized. The electron
polarization is then transferred to nucleon polarization through dipole-dipole
interaction between the electron and the nucleon.
By applying a suitable microwave signal, the desired spin state is populated.
NMR detectors are placed near the targets to monitor the target polarization. 
An average polarization of 80\% is expected for the NH$_3$ target.

The preparation of the SpinQuest is well underway and first physics data-taking
is expected to begin in September of 2019. Stay tuned.

\end{document}